# Characterization of nonlinear effects in edge filters

T. Amotchkina,[1*] M. Trubetskov,[1] E. Fedulova,[1,2]
K. Fritsch,[2] O. Pronin,[2] F. Krausz,[1,2] V. Pervak[2]

[1]*Max-Planck-Institut für Quantenoptik, Hans-Kopfermann-Str. 1, Garching 85748, Germany*
[2]*Ludwig-Maximilians-Universität München, Am Coulombwall 1, Garching 85748, Germany*
*\*Tatiana.Amotchkina@mpq.mpg.de*

**Abstract:** A specially designed and produced edge filter with pronounced nonlinear effects is carefully characterized. The nonlinear effects are estimated at the intensities close to the laser-induced damage.

## 1. Introduction

In the recent 20 years multilayer coatings have become a driving force in the field of ultrafast optics. The nonlinear effects manifest themselves under illumination of optical coatings by intense optical fields. In dielectric materials the field effectively acts on itself and a Kerr-type nonlinearity appears. This nonlinearity reveals itself as an intensity dependent addition to the refractive indices of the coating layers $n_2 I$ [1]. This effect can be used for the development of innovative laser components, for example, dielectric coatings with qualitatively and quantitatively predictable nonlinear properties. In order to use the potentials of dielectric coatings, their behaviour at intensities which are, on the one hand, high enough to activate nonlinearity and, on the other hand, lower than damage threshold, needs to be carefully studied. Furthermore, numerical values of the dielectric coatings nonlinear parameters are to be estimated.

In the present contribution, a specially tailored nonlinear edge filter (NEF) with an extremely steep spectral reflectance behaviour in the vicinity of the wavelength of 1030 nm is carefully characterized. The key values, intensity dependent reflectance $R(I)$ and transmittance $T(I)$ as well as the modulation depth $\Delta R(I)$ are measured and analyzed. Pronounced nonlinear behaviour at high intensities is demonstrated.

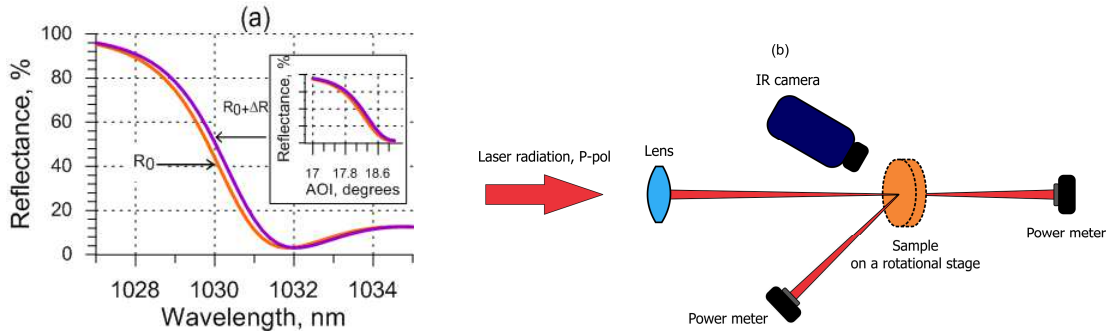

Fig. 1. (a) Sensitivity of reflectance of the NEF at high intensities; inset: reflectance of the NEF in the angular range (see the text for details); (b) Measurement setup for investigating the pre-damage behavior of the NEF.

## 2. Experimental results

The designed NEF consists of 69 layers and has a total physical thickness of 8.7 μm; $Nb_2O_5$ and $SiO_2$ were used as high and low index materials, respectively. The theoretical reflectance of the NEF is shown in Fig. 1(a). In the inset the angular dependence of the NEF's reflectance $R$ at a wavelength of 1030 nm is shown. This figure demonstrates high sensitivity of the NEF to any change in refractive indices, layer thicknesses and angle of incidence (AOI). The filter was deposited on fused silica substrates with thicknesses of 0.145 mm and 6.35 mm by magnetron sputtering at Helios plant from Leybold. The layer thicknesses were controlled using a well-calibrated time monitoring [2]. Good correspondence between theoretical and measured spectral characteristics of the sample was achieved.

The intensity dependent reflectance $R(I)$ and transmittance $T(I)$ were measured at the setup shown in Fig. 1(b). It includes an Yb:YAG thin disk regenerative amplifier with a repetition rate of 50 kHz [3]. The pulse intensity was adjusted by a half-wave plate followed by a polarization cube. The measurements were performed at intensities ranging from $3.3 \cdot 10^9$ W/cm² up to $63 \cdot 10^9$ W/cm². The latter value is about 5% below the damage threshold observed at $I \approx 66 \cdot 10^9$ W/cm² for this sample. The $R(I)$ and $T(I)$ values were calculated as follows:

$$R(I) = \frac{P_{ref}}{P}, \quad T(I) = \frac{P_{tr}}{P}, \quad I = \left(\frac{\pi}{2}\right)^{-3/2} \cdot \frac{P}{f_{rep} r^2 \tau}. \tag{1}$$

Here $P$, $P_{ref}$, $P_{tr}$ are the average incident, reflected and transmitted power measured via a power meter, respectively, $r$ is the beam radius, defined as transversal distance from the axis at which the intensity is $1/e^2$ times the peak intensity $I$, $f_{rep}$ is the repetition rate, $\tau$ is the pulse duration. In the performed measurements $f_{rep}$ was 50 kHz, $\tau$ was 1 ps, $r$ was 175 μm, $P$ was varying from 100 mW to 1.9 W. The sample temperature in the course of the measurements was monitored via a thermal imaging camera with an accuracy of ±2°C. The sample was placed on a rotational stage in order to adjust the AOI and to achieve different initial reflectance and transmittance values $R_0$ and $T_0$. Those values where adjusted at a lower laser output power corresponding to an intensity of $3.3 \cdot 10^9$ W/cm² at which no nonlinear effects in the sample are possible. The measured modulation depths $\Delta R(I) = R(I) - R_0$ at $R_0 = 50\%, 60\%, 70\%,$ and 80% are shown in Fig. 2(a) by red markers. The maximal $\Delta R$ is observed at $R_0 = 50\%$.

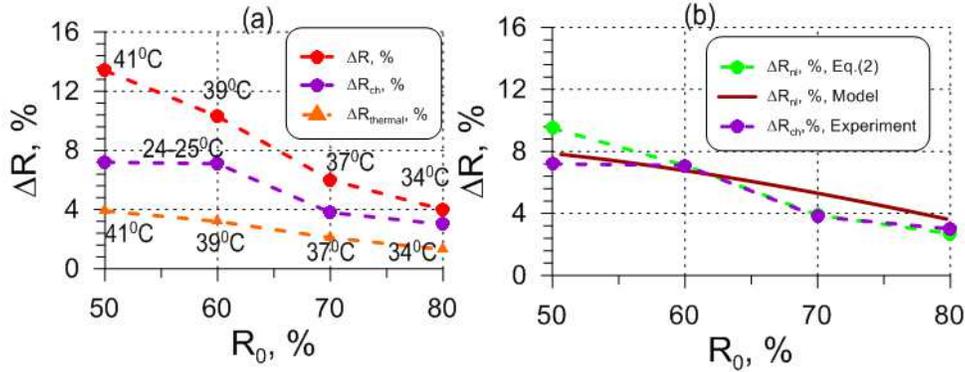

Fig. 2. (a): Measured modulation depths $\Delta R$ of the NEF, modulation depths measured with the optical chopper $\Delta R_{ch}$ and estimated thermal change of reflectance $\Delta R_{Thermal}$; (b) Correspondence between estimated nonlinear modulation depth $\Delta R_{nl}$ (Eq. (2)), modulation depth measured with the chopper and model estimation of $\Delta R_{nl}$ calculated under the assumption that $n_2 = 10^{-14}$ cm²/W.

The temperatures of the sample surface are indicated in Fig. 2(a) as well. The growth of the temperature can be explained by absorption, which is inevitable even in high quality multilayers. It is well known that heating of multilayer coatings shifts their spectral characteristics $R(\text{Temp})$ to longer wavelengths. Such thermal changes of the reflectance can be quantified as deviations between the reflectance at a reference wavelength of the sample at room temperature and reflectance of the heated sample at the same wavelength, $\Delta R_{Thermal} = R(\text{Temp}) - R_0$. In the case of the NEF, thermal changes lead to increasing reflectance as well. In order to distinguish between thermal and nonlinear effects, an optical chopper (10% duty cycle) was inserted into the laser beam after the lens (Fig. 1(b)). The measured modulation depths $\Delta R_{ch}$ are shown in Fig. 2 by violet markers. An increase of reflectance is still clearly observed and the rise of temperature is negligible. This can be considered as a clear indication for the presence of the nonlinear Kerr effect in the considered NEF.

For the purpose of cross checking, the thermal change $\Delta R_{Thermal}$ was estimated separately. The sample on the 6.35 mm thick fused silica substrate was heated during 1.5 hours in an oven at 125°C. Then its spectral transmittance $\Delta T_{Thermal}(\lambda)$ as well as surface temperatures during the cooling process were measured with the help of PerkinElmer Lambda 950 spectrophotometer and a portable thermal imaging camera, respectively (see Fig. 3(a)). The values of $\Delta R_{Thermal} = 100\% - \Delta T_{Thermal}$ corresponding to different $R_0$ are shown by the markers in Fig. 3(b).

The thermal changes of reflectance $\Delta R_{Thermal}$ observed in the course of the laser measurements (orange markers in Fig. 2(a)) were estimated using linear fits obtained from the spectral photometric measurements (Fig. 3(b)).

For a first rough estimation, we assume that the value of the nonlinear effect $\Delta R_{nl}$ can be estimated as:

$$\Delta R_{nl}(I) = \Delta R(I, P) - \Delta R_{Thermal}(\text{Temp}(P)) \tag{2}$$

The values $\Delta R_{nl}(I)$ (Eq. (2)) are presented in Fig. 2(b) by green markers. It is remarkable that these values are very close to the $\Delta R$ values obtained with the introduced chopper $\Delta R_{ch}$ (Fig. 2, violet markers).

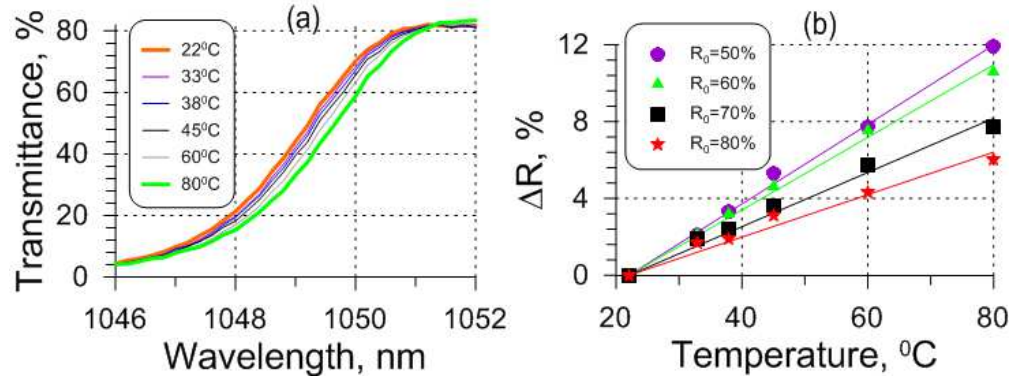

Fig. 3. (a): Transmittance of the NEF measured at temperatures indicated in the legend; (b): Thermal change of reflectance at different initial levels $R_0$; markers represent experimental data, solid lines indicate linear fits.

The solid curve in Fig. 2(b) represents the model of $\Delta R_{nl}$, calculated under the assumption that the nonlinear coefficient $n_2$ to the refractive index of $Nb_2O_5$ equals $10^{-14}$ cm$^2$/W. It is seen that model and experimental data are in a good agreement. Actually, these model calculations provide only a first crude approximation of $n_2$. This $n_2$ value is evidently overestimated because the dependence of the refractive index along the coating coordinate $n(I,z)$ was not taken into account. In order to provide a more accurate estimation of $n_2$, a special, sophisticated model and algorithm based on the numerical solution of the boundary-value problem derived from the system of Maxwell equations are required. It is remarkable, however, that the estimated $n_2$ value is of the same order of magnitude as $n_2(TiO_2)$ of $10^{-14}$ cm$^2$/W reported in the literature [4].

### 3. Conclusions

The novel dielectric multilayer filter exhibiting pronounced nonlinear increase of reflectance at high intensities has been produced and carefully characterized. A first approximation of the nonlinear addition $n_2$ for $Nb_2O_5$ thin film material has been found.

### 4. Acknowledgements

T. Amotchkina received funding from the European Union's Horizon 2020 research and innovation programme under the Marie Skłodowska-Curie agreement No 657596. The work of other authors was supported by the DFG Cluster of Excellence, "Munich Centre for Advanced Photonics," (http://www.munich-photonics.de). The authors are grateful to D. Ehberger, A. Ryabov, W. Schneider and P. Baum for providing and maintaining the laser source.